# Quantum Holography from Fermion Fields

Paola Zizzi

*Department of Brain and Behavioural Sciences, Pavia University, Piazza Botta, 11, 27100 Pavia, Italy*

e-mail: paola.zizzi@unipv.it

Phone: +39 3475300467

## Abstract

We demonstrate, in the context of Loop Quantum Gravity, the Quantum Holographic Principle, according to which the area of the boundary surface enclosing a region of space encodes a qubit per Planck unit. To this aim, we introduce fermion fields in the bulk, whose boundary surface is the two-dimensional sphere. The doubling of the fermionic degrees of freedom and the use of the Bogoliubov transformations lead to pairs of spin network's edges piercing the boundary surface with double punctures, giving rise to pixels of area encoding a qubit. The proof is also valid in the case of a fuzzy sphere.

## 1. Introduction

The holographic principle (HP), proposed by 't Hooft [1] and Susskind [2] states that the information entropy stored in a region of space of volume $V$ is encoded by the area $A$ of the boundary surface enclosing $V$, precisely one classical bit per unit of Planck area.

HP is based on the thermodynamics of black holes, and is in a sense a generalization of the Bekenstein bound [3]. In particular, the original version of 't Hooft provides a spherical entropy bound, followed by the covariant entropy bound of Bousso [4].

The holographic principle has been applied to the study of black hole thermodynamics and quantum gravity, and has been exactly formulated in the AdS/CFT correspondence [5].

Susskind [2] has applied the HP mostly to string theories, while Rovelli [6] and others [7] [8] have recovered the HP in the context of Loop Quantum Gravity (LQG) [9].

In the present work, we follow the LQG approach, instead of that of string theory.

In LQG the gravitational quantum excitations are the spin networks [10], originally introduced by Penrose [11], which account for the black hole entropy in accordance with the HP. As the states which dominate the counting of the degrees of freedom correspond to punctures of spin $j=1/2$, each pixel of area encodes a classical bit of information.

In this paper, we consider the quantum version [12] [13] [14] of the HP, namely, the quantum holographic principle (QHP), or "quantum holography", according to which the information encoded by the boundary surface is one qubit (per unit of Planck area) instead of a classical bit.

Also, we will focus on 't Hoof's spherical entropy bound, and find that in the QHP, the number of qubits grows logarithmically with the surface area, differently from the classical case where the number of bits is directly proportional to the surface area.

To demonstrate the QHP in the context of LQG, it will be necessary to introduce a dissipative quantum field theory (DQFT) [15-20] in the bulk. In particular, we will consider fermionic DQFT. The latter, as well as the associated Bogoliubov transformations [21], were first introduced by Takahashi and Umezawa [22] and then revisited in [23].

We will take into account the following important features of QFT.

First, QFT has infinite degrees of freedom, and this allows the existence of infinite unitarily inequivalent representations (u.i.r.) of the CCR, unlike QM, which has a finite number of degrees of freedom, so that, by the Stone-von Neumann's theorem [24] [25] in QM the representations of the CCR are all unitarily equivalent to each other.

Second, in the case of a dissipative QFT, one has to double the degrees of freedom in the environment.

In this paper, by doubling the fermionic degrees of freedom of the fermionic DQFT, and using the Bogoliubov transformations for fermions, we show that the spin network's edge labelled by $j = +1/2$, which pierces the boundary surface from the inside (the bulk) has a dual with opposite spin $j = -1/2$, which pierces the surface from the outside (the environment). The "double puncture" gives rise to a pixel of area that encodes a qubit.

We also show that, if instead of the ordinary sphere we consider the fuzzy sphere [26] in the *N=2* representation of *SU(2)*, the double puncture consists of two cells, one cell encodes the bit 0 and the other cell encodes the bit 1, so that the *N=2* fuzzy sphere as a whole encodes a qubit.

In this case the doubling of the degrees of freedom is already inherent in the two cells and the fermion field is reduced directly to a "peak" of quantum information.

In both cases, however, the quantum information that is encoded by the boundary surface is only the "hidden quantum information" [27] released by the fermion fields. The topic of fermions on the fuzzy sphere is a relatively old research topic in mathematical physics, dating back to the 1990s, with an article by Grosse and Prešnajder [28] analyzing the fuzzy analogue of spinor bundles over the sphere. The paper is organized as follows.

In Sect. 2, we give the new entropy bound for the QHP. We will also briefly review spin networks, their application to black hole entropy, and their role in the QHP. Finally, we will illustrate the reasons why the introduction of fermionic DQFT in the bulk is necessary to demonstrate the QHP.

In Sect. 3, we double the fermionic degrees of freedom and perform the Bogoliubov transformations for the fermions: this leads to pairs of spin network edges with opposite spins, which pierce the surface with double punctures and give rise to pixels encoding a qubit.

In Sect. 4, we shortly review the fuzzy sphere and its relation with the QHP, and show how the fermion field's excitation at a given point in ordinary space corresponds to a "spike" of quantum information in a non-commutative space. Section 5 is devoted to the conclusions.

## 2. Why QFT for quantum holography? The logical thread of the problem

In this Section, we will briefly review the (classical) Holographic Principle (and we will also suggest a new entropy bound in the quantum case), as well as spin networks and their application to black holes entropy in the context of LQG. At the end, we will discuss the reasons why it is necessary to start from a fermionic DQFT in order to demonstrate the QHP in the context of LQG.

## 2.1 A new entropy bound for QHP

The spherical entropy bound introduced by 't Hooft [1] is the extension of the Bekenstein bound [3] to whichever region of space of volume *V*:

$$S \le \frac{A}{4L_p^2} \tag{2.1}$$

Where $S$ is the entropy, $A$ is the area of the surface bounding $V$, and $L_p$ is the Planck length. For black holes, Eq. (2.1) is saturated, giving back the Bekenstein bound.

The HP states that the information entropy inside the bulk of a region of space can be by described at the boundary, by a number $n$ of Boolean degrees of freedom which should not be larger than one bit per Planck area.

In a discrete theory of $n$ spins, the number of microstates is $2^n$, which is given by:

$$2^n = \exp S \tag{2.2}$$

and from Eq. (2.1) it follows:

$$n \leq \frac{A}{4 \ln 2 L_p^2} \qquad . \tag{2.3}$$

In the QHP the number of microstates is still $2^n$, but now it is:

$$n = 2^N \tag{2.4}$$

where $N$ is the number of qubits, thus we have:

$$2^N \leq \frac{A}{4 \ln 2 L_p^2} \qquad . \tag{2.5}$$

From Eq. (2.5) it follows:

$$N = \frac{1}{\ln 2} \ln \left( \frac{A}{4 \ln 2 L_p^2} \right) \qquad . \tag{2.6}$$

Eq. (2.6) shows that the total amount of quantum information (the number $N$ of qubits) encoded by a black hole event horizon is proportional to the logarithm of the horizon area.

Of course $N \propto \ln A$ grows much more slowly than $n \propto A$. This leads us to suppose that the algorithm to be used for the simulation of a black hole must be a quantum algorithm which is exponentially faster, that is with running time $O(2^N)$, than a classical one with running time $O(n)$.

## 2.2 Spin networks and black hole entropy

Spin network were applied [10] to LQG, where they are eigenstates of the area and volume-operators [29] [30].

When a 2-dimensional surface is punctured transversally by one network's edge, the puncture gives rise to an area:

$$A_1 \propto L_P^2 \sqrt{j(j+1)} \qquad . \tag{2.7}$$

If the surface is punctured in $n$ points, the area is:

$$A_n \propto L_P^2 \sum_n \sqrt{j_n(j_n+1)} \qquad . \tag{2.8}$$

The edges of the spin network, as they pierce the horizon of the black hole, they excite the degrees of freedom of curvature on the surface [6-8], which account for the black hole entropy:

$$S_{BH} \propto \frac{A}{4 L_P^2 \gamma} \tag{2.9}$$

Where $S_{BH}$ is the black hole entropy, $A$ is the horizon's area, $L_P$ is the Planck length and $\gamma$ is the Immirzi parameter [31].

The states predominating the counting of degrees of freedom correspond to punctures labelled by $j = 1/2$. Then, the origin of black hole entropy seems strictly related to the Wheeler's "it from Bit proposal [32]. A pixel can be either "true" ("on" ) $\equiv 1$ or "false" ("off" ) $\equiv 0$, where we take the convention that the pixel is "true" when the puncture is made by the edge of a spin network (in the $N = 2$ representation of *SU(2)*) in the state $|+1/2\rangle$ and that the pixel is "false " when the spin network 's edge is in the state $|-1/2\rangle$.

In the QHP the pixel can be "true" and "false" at the same time, that is the pixel will encode a qubit, rather than a bit. It must be said that the underlying logic in this case is no longer Boolean, but paraconsistent [33] as the non-contradiction principle $A \wedge \neg A = F$ (where $A$ is a proposition, $\neg A$ is it negation and $F$ is the false) is violated.

## 2.4 From fermionic QFT to QHP

Why derive the QHP from a fermionic DQFT and not assume it directly in QM? In this section we will present, quite informally, the reasons why it is not possible to formulate the QHP directly in QM, unlike the standard (classical) holographic principle.

Quantum Field Theory itself is not a quantum-computational theory, differently from Quantum Mechanics. However, it was shown in [27] that QFT encodes hidden quantum information, which can be revealed through a suitable reduction mechanism [34] [27] directly down to Quantum Information (QI). It would be worth applying this intrinsic property of QFT to demonstrate the QHP, where the information encoded by the boundary surface is quantum that is, given in terms of qubits. This would require a reduction of the infinite degrees of freedom of QFT to a finite number.

The HP states that all the information enclosed in a region of space of volume $V$ is encoded on the surface of area $A$ surrounding $V$, precisely, one bit of classical information for each pixel of Planck area. In LQG such encoding is realized by the punctures of spin network's edges [10] on the boundary surface, more precisely each puncture gives rise to a pixel of Planck area encoding one bit of information.

The QHP states that each pixel of area of the surface surrounding $V$ encodes one qubit. In the QHP in the context of LQG [13], then, one should expect that the boundary surface $\partial\Sigma$ (where $\Sigma$ is a 3-dimensional hypersurface) is punctured simultaneously from the inside and from the outside. The spin network's edge labelled by $+1/2$, which punctures the boundary surface $\partial\Sigma$ at the point $P_0 = (x_0, y_0)$ from inside, gives rise to a pixel of Planck area encoding the bit $|1\rangle$ and the symmetric spin network's edge labelled by $-1/2$, which punctures $\partial\Sigma$ at the same point from the outside, supplies the same pixel with the other bit $|0\rangle$. See Fig.1

**Fig.1**

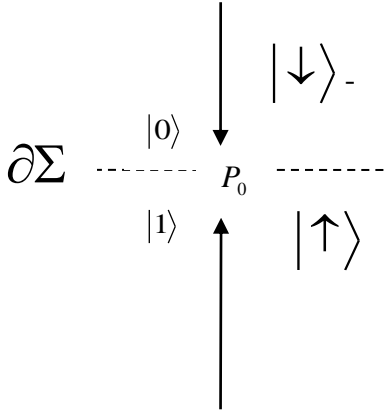

**A Double Puncture**: A pair of spin edges "up" $\left|\uparrow\right\rangle$ and "down" $\left|\downarrow\right\rangle$ pierce the boundary $\partial\Sigma$ at the same point $P_0$ simultaneously and the pixel encodes both bits $\left|0\right\rangle$ and $\left|1\right\rangle$.

However, LQG describes spin networks as the quantum states $\left|j,m\right\rangle$ (the eigenstate of angular momentum in the spherical representation) of a discrete geometry of the bulk, that is, the spin network's edges can puncture the boundary surface only from inside. To avoid this restriction, one might be tempted to double the quantum states $\left|j,m\right\rangle$ as in the case of a (damped) quantum harmonic oscillator (QHO). However, two main problems arise.

First, in the case of fermions, the creation and annihilation operators cannot be defined as usual in bosonic QHO, as they have no analogue in the classical theory because of the spin quantum numbers, so that they must be defined directly in QFT, from the Dirac equation, as was showed in [22] [23].

Second, the time evolution would take the CCR out from the original Hilbert space, and in QM one cannot fix this problem because of the Stone-von Neumann theorem [24] [25], which states that in QM all the representations of the CCR are unitarily equivalent.

Instead, QFT has an infinite number of degrees of freedom and the Stone-von Neumann's theorem does not hold, so that QFT admits infinite u.i.r. of the CCR. Then, the only way out is to start directly with a fermionic DQFT on the hypersurface $\Sigma$. This should lead us to the QHP through the following 8 steps:

1) Consider a fermionic quantum field theory on $\Sigma$.

2) Double the degrees of freedom of the fermion field in the environment E.

3) Consider the discrete geometry of LQG (spin network's edges that puncture $\partial\Sigma$). For simplicity, we will take $\partial\Sigma$ to be the ordinary sphere $S^2$.

4) Assume that at each puncture it corresponds an excitation of the fermion field (i.e. a half-integer spin particle).

5) Because of the doubling of the degrees of freedom performed on the fermion field, at each puncture, a fermionic particle on $\Sigma$ has its double in E.

6) From 4) and 5) it follows that, through a suitable projection of the particle living in the environment E, at each puncture, it is possible to associate a spin network's edge in E, which is "dual" to the original one in $\Sigma$.

7) A Bogoliubov transformation on the fermionic operators allows us to interpret the "dual" spin network's edge as the reflected image of the original one.

8) All the above mathematical procedure performed on the ordinary sphere $S^2$ is more straightforward when we consider the fuzzy sphere [26] $S_F^2$ instead of $S^2$. In this case the background space is itself quantum, and this induces a reduction of the infinite degrees of freedom of the fermion field to a finite number. What remains of the original fermion field at a point (the latter becoming a cell) is what we will call a "spike" of quantum information.

The geometrical quantization in terms of spin networks on the ordinary sphere $S^2$ discussed above reduces the infinite degrees of freedom to a finite number, but there is the need of doubling the degrees of freedom in order to achieve the QHP. Instead, the Lorentz invariant "regularization" of the fermion field on the non-commutative "lattice", where the sites are the cells of the fuzzy sphere $S_F^2$ (corresponding to the punctures of the spin network's edges on the ordinary sphere) leads automatically to the QHP.

## 3. Doubling the fermionic degrees of freedom leads to double punctures

In this Section, we will show how doubling the fermionic degrees of freedom leads, through Bogoliubov transformations, and through appropriate projections of the field excitations at a point on the boundary surface, to the appearance of pairs of opposite spin network edges, which pierce the boundary surface with a double puncture.

### 3.1 Bogoliubov transformations for fermions

Fermionic QFT, described by the fermionic creation and destruction operators $\left(f_k, f_k^\dagger\right)$ lives then on the three-dimensional hypersurface $\Sigma$. Let us suppose for simplicity that $\Sigma$ is a three-dimensional ball that encloses a portion of space of volume $V$. We denote by $\partial\Sigma$ the 2-dimensional boundary of $\Sigma$, that is the sphere $S^2$. Following the mathematical setting in [22] [23], we double the degrees of freedom of the original fermionic QFT, and we get two fermionic quantum field theories, one on $\Sigma$, and its double in the environment E.

Given two sets of fermionic annihilation and creation operators $\left(f_{1k}, f_{1k}^\dagger\right)$ and $\left(f_{2k}, f_{2k}^\dagger\right)$, satisfying the anti-commutation relations: $\left\{f_{ik}, f_{jk'}^\dagger\right\} = \delta_{ij}\delta_{kk'}$ $(i, j = 1,2)$, the Bogoliubov transformations for fermions are:

$$f_{1k} \to f_{1k}(\theta) = U^{-1} f_{1k} U = f_{1k} \cos\theta_k - f^\dagger{}_{2k} \sin\theta_k \tag{3.1}$$

$$f_{2k} \to f_{21k}(\theta) = U^{-1} f_{2k} U = f_{2k} \cos\theta_k + f^\dagger{}_{1k} \sin\theta_k$$

where $U$ is the unitary operator:

$$U(\theta) = \exp\left(iG(\vartheta)\right) \tag{3.2}$$

with generator:

$$G(\theta) = i \sum_k \theta_k \left( f^{\dagger}_{1k} f^{\dagger}_{2k} - f_{2k} f_{1k} \right). \tag{3.3}$$

We remind, however, that we are looking for a spin network's edge labelled by $j = +1/2$ which punctures the boundary at a given point, let us call it $P_0 = (x_0, y_0)$ from inside (the bulk), and of a "dual" spin network's edge labelled by $j = -1/2$ which punctures the boundary at the same point from outside (the environment). Then, we should consider the fermion field on $\Sigma$, $\Psi(\bar{x})_{\Sigma}$ at point $\bar{x}_0 = (x_0, y_0, z_0) \in \Sigma$, and its "double" $\Psi(\bar{x})_E$ at point $\bar{x}_0' = (x_0, y_0, -z_0) \in E$. As it is well known, in QFT a particle (in our case a fermion) is an excited state at a point of the field. If we consider the point $\bar{x}_0 \in \Sigma$, the probability amplitude of having the particle at $\bar{x}_0$ is given by $\langle \Psi(\bar{x}) \| \Psi(\bar{x}_0) \rangle_E$, and the probability density is given by:

$$p_0 = \left| \langle \Psi(\bar{x}) \| \Psi(\bar{x}_0) \rangle_{\Sigma} \right|^2. \tag{3.4}$$

In the same way, for the "double", the probability amplitude of having the particle at $\bar{x}_0' \in E$ is given by $\langle \Psi(\bar{x}) \| \Psi(\bar{x}_0') \rangle_E$, and the probability density is given by

$$p_0' = \left| \langle \Psi(\bar{x}) \| \Psi(\bar{x}_0') \rangle_E \right|^2. \tag{3.5}$$

Bijective mapping between qubits and fermion states helps determine the number of qubits needed to simulate a fermionic QFT.

To achieve this, the fermion field should be regularized on a lattice. However this procedure, in particular regularization on lattice QCD, shows some well known difficulties like, for example, fermion doubling, which is strictly related to chirality [35] (incidentally, the term "fermion doubling" should not be confused with the "doubling of fermionic degrees of freedom" used in this work).

We suggest that a way out is to take the finite set of punctures in LQG as the most natural discrete space with regard to our particular case. The only new issue is to attach, to each puncture $P_i = (x_i, y_i, z_i = 0) \in \partial \Sigma$ a probability density:

$$p_i = \left| \langle \Psi(\bar{x}) \| \Psi(\bar{x}_i) \rangle_{\Sigma} \right|^2 = \left| \langle \Psi(\bar{x}) \| \Psi(\bar{x}_i') \rangle_E \right|^2. \tag{3.6}$$

This will become clearer in Sect. 4, where we will consider a fuzzy sphere instead of the ordinary sphere.

## 3.2 Projections of fermions

The first problem one faces is how to reduce the infinite degrees of freedom carried by the quantum field $|\Psi(\bar{x}_0)\rangle_{\Sigma}$ (and its double $|\Psi(\bar{x}_0')\rangle_E$) to the finite degrees of freedom of the quantum state $|\uparrow\rangle$ (and $|\downarrow\rangle$).

The original field $\Psi(\bar{x})_{\Sigma}$ and its "double" $\Psi(\bar{x})_E$ form a closed system in the total Hilbert space. The two fermions $|\Psi(\bar{x}_0)\rangle_{\Sigma}$ and $|\Psi(\bar{x}_0')\rangle_E$ are the mirror image of each other for a reflection $z \to -z$ with respect to the plane $x, y$. Let us write the Dirac field on $\Sigma$ in terms of two 2-component fermions:

$$\left| \Psi(\bar{x}) \right\rangle_{\Sigma} = \begin{pmatrix} \varphi(\bar{x}) \\ \eta\ (\bar{x}) \end{pmatrix}_{\Sigma},$$ (3.7)

and its "double" as well:

$$\left| \Psi(\bar{x}^{'}) \right\rangle_{E} = \begin{pmatrix} \varphi(\bar{x}^{'}) \\ \eta\ (\bar{x}^{'}) \end{pmatrix}_{E}.$$ (3.8)

Through a projection $z = 0$ on $\partial\Sigma$, let us call it $\perp_{\partial\Sigma}$, we have:

$$\perp_{\partial\Sigma} \left| \Psi(\bar{x}_0) \right\rangle_{\Sigma} = \perp_{\partial\Sigma} \left| \Psi(\bar{x}_0^{'}) \right\rangle_{E} = \left| \Psi(x_0, y_0) \right\rangle_{\partial\Sigma} = \begin{pmatrix} \varphi(x_0, y_0) \\ \eta\ (x_0, y_0)_{\partial\Sigma} \end{pmatrix}.$$ (3.9)

Then, the QHP (differently from the HP) has an intrinsic probabilistic nature: the two spin network's edges $j = +1/2$ and $j = -1/2$ will puncture simultaneously the boundary at point $P_0 = (x_0, y_0)$ with probability given by Eq. (3.4).

The Dirac fermion field $\Psi(\bar{x})$ at a fixed point $\bar{x}_0 = (x_0, y_0, z_0) \in \Sigma$ can be written as the quantum state:

$$\left| \Psi \right\rangle = \begin{pmatrix} 0 \\ 1 \\ 1 \\ 0 \end{pmatrix} = \left| 1 \right\rangle \oplus \left| 0 \right\rangle.$$ (3.10)

Here, the fermionic operators $\left( f_k, f_k^{\dagger} \right)$ are given in the representation of generalized projectors:

$$f = \left| 0 \right\rangle \left\langle 1 \right|\ ,\ f^{\dagger} = \left| 1 \right\rangle \left\langle 0 \right|.$$ (3.11)

Now, let us project the quantum state $\left| \Psi \right\rangle_2$ in the environment E as:

$$f_2 \left| \Psi \right\rangle_2 = \begin{pmatrix} 0 \\ 0 \\ 1 \\ 0 \end{pmatrix} = \left| \downarrow \right\rangle_E \qquad\qquad f_2^{\dagger} \left| \Psi \right\rangle_2 = \begin{pmatrix} O \\ 1 \\ O \\ O \end{pmatrix} = \left| \uparrow \right\rangle_E$$ (3.12)

and the quantum state $\left| \Psi \right\rangle_1$ in $\Sigma$ as:

$$f_1 \left| \Psi \right\rangle_1 = \begin{pmatrix} 0 \\ 0 \\ 1 \\ 0 \end{pmatrix} = \left| \downarrow \right\rangle_B \qquad\qquad f_1^{\dagger} \left| \Psi \right\rangle_1 = \begin{pmatrix} 0 \\ 1 \\ 0 \\ 0 \end{pmatrix} = \left| \uparrow \right\rangle_B$$ (3.13)

where the subscripts $E$ and $B$ stand for "environment" and "bulk" respectively.
As from the bulk only spin network's edges with spin "up" puncture the boundary transversally, we retain only the projection $f_1^{\dagger} \left| \Psi \right\rangle_1$ of the fermion field on $\Sigma$. The inverse holds for the environment: only spin network's edges with spin "down" puncture the boundary transversally, so that we will retain only the projection $f_2 \left| \Psi \right\rangle_2$ for the fermion field in $E$.

### 3.3 Rotations of the spinors

The relation between $f_1^\dagger$ and $f_2$ is given by a piece of the Bogoliubov transformations for fermions [22] [23]:

$$f_{2k} \to f_{2k}(\vartheta) = f_{2k} \cos\theta_k + f_{1k}^\dagger \sin\theta_k \quad . \tag{3.14}$$

For $\vartheta = \pi/2$, it holds $f_{2k} \to f_{1k}^\dagger$, which corresponds to a counter clockwise rotation of $180°$ around the x-axis on the spin $j = -1/2$.

The spin rotation operator of an angle $\vartheta$ about the x-axis is:

$$D_x(\vartheta) = \exp\left(-i\frac{\vartheta}{2}\sigma_x\right) = \begin{pmatrix} \cos(\vartheta/2) & i\sin(\vartheta/2) \\ i\sin(\vartheta/2) & \cos(\vartheta/2) \end{pmatrix} \quad . \tag{3.15}$$

Through the action of the matrix $\exp\left(-i\dfrac{\vartheta}{2}\sigma_x\right)$ on the spinor, the flagpole is rotated by the angle $\vartheta$ about the x-axis.

For $\vartheta = 180°$ it is:

$$D_x(\pi) = \exp\left(-i\frac{\pi}{2}\sigma_x\right) = i\begin{pmatrix} 0 & 1 \\ 1 & 0 \end{pmatrix} = i\sigma_x \quad . \tag{3.16}$$

The action of $D_x(\pi)$ on the spin "down" state $|\downarrow\rangle_E = \begin{pmatrix} 1 \\ 0 \end{pmatrix}_E$ is (up to a global phase):

$$D_x(\pi)|\downarrow\rangle_E = |\uparrow\rangle_B \tag{3.17}$$

where $|\uparrow\rangle_B = \begin{pmatrix} 0 \\ 1 \end{pmatrix}_B$ is the spin "up" state.

In the same way, the relation between $f_2^\dagger$ and $f_1$ is given by the other piece of the Bogoliubov transformations:

$$f_{1k} \to f_{1k}(\theta) = f_{1k} \cos\theta_k - f_{2k}^\dagger \sin\theta_k \quad . \tag{3.18}$$

For $\vartheta = \pi/2$, it holds $f_{1k} \to -f_{2k}^\dagger$, that corresponds to a clockwise rotation of an angle of $180°$ around the x-axis of the spin $j = +1/2$:

$$D_x(\pi)|\uparrow\rangle_B = |\downarrow\rangle_E \quad . \tag{3.19}$$

The projection of a generic $j = 1/2$ spin network state in the bulk $|j = 1/2, m\rangle_B$ along the z-axis gives: $|j = 1/2, m = +1/2\rangle_B \to \begin{pmatrix} 1 \\ 0 \end{pmatrix}_B$ , $|j = 1/2, m = -1/2\rangle_B \to \begin{pmatrix} 0 \\ 1 \end{pmatrix}_B$, which are the basis states $|0\rangle_B, |1\rangle_B$ respectively.

The two spin network's edges $|\downarrow\rangle_E$ and $|\uparrow\rangle_B$ now puncture $\partial\Sigma$ at the same point, from outside and from inside respectively. We have realized the quantum holographic principle starting from a fermionic DQFT on the hypersurface, doubling the degrees of freedom, performing the Bogoliubov transformations, considering both the original field and its double at the same fixed point, made a generalized projection to recover the spin network's edges states in the bulk and in the environment, which are flipped into each other as a consequence of the Bogoliubov transformations on the fields. By making appropriate projections on Dirac fermions, we are able to retrieve the edges of the spin network labelled +1/2 (-1/2) piercing the boundary from the inside (the bulk) and from the outside (the environment) respectively. In this way the quantum holographic principle is recovered, in fact such a two-ways perforation gives rise to a pixel of area that is "on" and "off" at the same time, and therefore encodes a qubit.

## 4. Fermions and the fuzzy sphere

In this Section we will briefly review the fuzzy sphere [26], and its relation with the QHP, in the context of LQG. Then, we will show how fermion fields reduce to quantum information "spikes" on the fuzzy sphere in the fundamental (*N=2*) representation of *SU(2)*.

### 4.1 The fuzzy sphere and the QHP

The fuzzy sphere [26] is constructed by quantizing the coordinates $x_i$ $(i=1,2,3)$ of the ordinary sphere, that is, by replacing the $x_i$ by the non-commutative coordinates $X_i$:

$$x_i \rightarrow X_i = kJ_i \qquad\qquad (i=1,2,3) \qquad\qquad (4.1)$$

where the $J_i$ form the *N*-dimensional irreducible representation of the algebra of *SU(2)*, and $k$ is a parameter called the non-commutativity parameter:

$$k = \frac{r}{\sqrt{N^2 - 1}} \ . \qquad\qquad (4.2)$$

The dimension *N* of the irreducible representation of *SU(2)* is equal to the number of elementary cells of the fuzzy sphere.

In the case $N=2$ the non-commutative coordinates $X_i$ are given in terms of the Pauli matrices $\sigma_i$:

$$X_i = k\sigma_i \qquad\qquad (i=1,2,3) \quad . \qquad\qquad (4.3)$$

and Eq. (4.2) becomes:

$$k = \frac{r}{\sqrt{3}} \qquad . \qquad\qquad (4.4)$$

From Eq. (4.2) it follows that $k \rightarrow 0$ for $N \rightarrow \infty$, and one recovers the ordinary sphere $S^2$.

The area of an elementary cell of the fuzzy sphere in the *N*-dim. irreducible representation of *SU(2)* is [13]:

$$A_N^{\ EC} = \frac{4\pi r^2}{\sqrt{N^2 - 1}} \quad . \qquad\qquad (4.5)$$

For $N \rightarrow \infty$, $A_N^{\ EC} \rightarrow 0$, that is, the elementary cell reduces to a point, and the fuzzy sphere tends to the ordinary sphere.

In the particular case of $N=2$, there are two elementary cells, each one of area:

$$A_2^{\ EC} = \frac{4\pi}{\sqrt{3}} \qquad\qquad (r=1) \ . \qquad\qquad (4.6)$$

The QHP is strictly related to 2-dimensional non-commutative geometry. In fact, for the case of one qubit, it was shown [36] that the geometrical representation of the qubit state space (the Bloch sphere) is in a one-to-one correspondence with the fuzzy sphere in the fundamental ($N=2$) representation. In [36] and [13] the mathematical setting was based on C*-algebras [37] and the application of the non-commutative version of the Gelfand-Naimark theorem [38].

Each one of the two elementary cells is a *SU(2)* doublet $\xi_\alpha$ $(\alpha = 1,2)$:

$$\xi_1 = \begin{pmatrix} 1 \\ 0 \end{pmatrix} \qquad \xi_2 = \begin{pmatrix} 0 \\ 1 \end{pmatrix} \quad . \tag{4.7}$$

The area of each of the two cells is [34]:

$$A^i{}_{N=2}{}^{EC} = \frac{4\pi}{\sqrt{3}} \, p(\xi_\alpha) \tag{4.8}$$

where $p(\xi_\alpha)$ is the probability to find $\xi_\alpha$ in one of the two cells, and the fuzzy sphere encodes in total one qubit. For a cat state encoded by the $N = 2$ fuzzy sphere it holds:

$$p(\xi_1) = p(\xi_2) = \frac{1}{\sqrt{2}}, \quad p(\xi_1)^2 + p(\xi_2)^2 = 1. \tag{4.9}$$

## 4.2 From fermion fields to quantum information "spikes"

The fermion field $\Psi_\alpha(\bar{x})$, where $\alpha$ is the spin index, on the standard sphere $S^2$, has an excitation at each puncture $\bar{x} = \bar{x}_0$ made by a spin network's edge labelled by $j = 1/2$, that is, a fermionic particle $\Psi_\alpha(\bar{x}_0)$.

Once the fermion field $\Psi_\alpha(\bar{x})$ is considered on the fuzzy sphere $S_F^2$, it undergoes a reduction of the infinite degrees of freedom to a finite number due to the non-commutative character of the variables $X_i = k J_i$:

$$\Psi_\alpha(\bar{x}) \to \Psi_\alpha(X_i) \qquad (i = 1,2,3) \, . \tag{4.10}$$

For $N = 2$, we define:

$$\Psi_\alpha(X_i) \equiv k \sigma_i \zeta_\alpha \qquad (\alpha = 1,2) \quad , \; (i = 1,2,3) \quad . \tag{4.11}$$

We will call the object in Eq. (4.11) a "spike", more precisely, a quantum information spike (QIS). It is the analogue, on a cell of $S_F^2$, of a fermion at $\bar{x} = \bar{x}_0$ on $S^2$, although the analogy is purely geometrical, and not physical. Put simply, the "spike" consists of the quantum information hidden in the quantum field at a point, which is revealed when the field is deprived of its dependence on ordinary spatial coordinates.

Quantum fields have infinite degrees of freedom due to their dependence on space-time variables (in our case only space variables since we are considering a hypersurface). However, if the space considered is a quantum space (in the sense that the spatial coordinates are not commutative) the number of degrees of freedom is no longer infinite, but is reduced to a finite number.

This becomes evident when we consider the simplest case, that of a fuzzy sphere in the fundamental representation ($N = 2$) of *SU(2)*. As we have already mentioned, there are only two points, the North Pole and the South Pole that have been smeared in two cells. All the other points no longer exist, and the original field no longer depends on infinite variables, but only on two, which are not even really two points, but two cells, each of which encodes a bit. Then, the fermion field has shrunk to a QIS given by Eq. (4.11). If we call $\eta$ the number of degrees of freedom that remain when the field is considered on a fuzzy sphere, we see that $\eta = N$, which is the number of cells of the fuzzy sphere in the $N$-dimensional irreducible representation of *SU(2)*.

For $k = 0$ one gets:

$$\Psi_\alpha (X_i) = 0, \ N \to \infty, \ \eta \to \infty, \ S_F^2 \to S^2. \tag{4.12}$$

The greater the dimension $N$ of the representation of SU (2), the greater the number of elementary cells, and for $N$ tending to infinity, the number of cells tends to infinity, and their area becomes smaller and smaller tending to zero, i.e. a point:

$$A_N^{\ EC} = \frac{4\pi r^2}{\sqrt{N^2 - 1}} \to 0 \qquad \text{for } N \to \infty \ . \tag{4.13}$$

This means that the larger the $N$-dimensional representation of *SU(2)*, the smaller the number of reduced degrees of freedom of the quantum field.

Conversely, the lower the $N$-dim. representation of *SU(2)*, the greater the fuzziness, the less degrees of freedom left. The minimum number of the latter is $\eta = 2$, corresponding to the fundamental representation.

The maximum reduction occurs for the fundamental representation $N = 2$. Moreover, only for $N = 2$ the "spikes" obey the anti-commutation relations which resemble those of the fermionic field, because in this case the non-commutative coordinates $X_i$ are proportional to the Pauli matrices, which satisfy the anti-commutation relations $\{\sigma_i, \sigma_j\} = 2\delta_{ij} I_2$ $(i, j = 1,2,3)$, where $I_2$ is the 2-dimensional identity matrix.

Furthermore, the number $N$ of cells is related to the number $n$ of qubits encoded by the fuzzy sphere by the relation [13]:

$$N = 2^n. \tag{4.14}$$

Eq. (4.14) is a constraint that was imposed [13] on $N$ such that the fuzzy sphere with $N$ cells can encode $n$ qubits. The computational state of a quantum computer with $n$ qubits can be geometrically viewed as a fuzzy sphere with $N = 2^n$ cells.

For $N \to \infty$, $n \to \infty$, that is, the amount of quantum information encoded in ordinary space, and consequently in a quantum field is infinite although hidden. It reveals itself only when the infinite degrees of freedom are reduced to a finite number. We recall the definitions:

$$N = 2^n, \ k = \frac{1}{\sqrt{N^2 - 1}} \ \text{(for } r = 1) \ , \ X_i = kJ_i \ (i = 1,2,3) \tag{4.15}$$

and give scheme A below.

### Scheme A

For: $2 \le N < \infty$ , $\qquad k \ne 0$ $\qquad\qquad\qquad S^2 \to S_F^2$

$\qquad 1 \le n < \infty$ $\qquad\qquad\qquad\qquad\qquad\qquad \Psi(x_i) \to \Psi(X_i)$

$\qquad\qquad\qquad\qquad\qquad\qquad\qquad\qquad\qquad \eta = \infty \to \eta = N$

For: $N \to \infty$ $\qquad\qquad k = 0$ $\qquad\qquad\qquad\qquad S_F^2 \to S^2$

$\qquad n \to \infty$ $\qquad\qquad\qquad\qquad\qquad\qquad\qquad \Psi(X_i) \to \Psi(x_i)$

$\qquad\qquad\qquad\qquad\qquad\qquad\qquad\qquad\qquad \eta = N \to \eta = \infty$

The Hermitian conjugate of $\Psi_\alpha(X_i)$ in Eq. (4.11) is:

$$\Psi_\alpha^{\ '}(X_i) \equiv k\sigma_i \zeta_\alpha^{\ '} \quad . \tag{4.16}$$

The QIS $\Psi_\alpha(X_i)$ is normalized:

$$\langle \Psi_\alpha(X_i) \| \Psi_\alpha(X_i) \rangle = 1 \tag{4.17}$$

as it holds: $k^2 = \dfrac{1}{3}$, $\sum_{i=1}^{3}\sigma_i^2 = 3$, $\langle \zeta_\alpha \| \zeta_\alpha \rangle = 1$.

For $N = 2$, the $\Psi_\alpha(X_i)$ satisfy the anti-commutation relations:

$$\left\{ \Psi_\alpha(X_i), \Psi_\beta^{\ '}(X_j) \right\} = \frac{2}{3}\delta_{ij}I_2\delta_{\alpha\beta} \quad (i,j = 1,2,3) \ \ \alpha,\beta = 1,2 \quad . \tag{4.18}$$

The excitation $\Psi_\alpha(\bar{x}_0)$ of the fermion field (i.e., a fermionic particle) at a puncture $\bar{x} = \bar{x}_0$ on the ordinary sphere $S^2$, becomes, on the fuzzy sphere $S_F^2$:

$$\Psi_\alpha(\bar{x}_0) \rightarrow I_2\zeta_\alpha \qquad (\alpha = 1,2) \quad . \tag{4.19}$$

The probability amplitude $\langle \psi_\alpha(\bar{x}) \| \psi(\bar{x}_0) \rangle$ to find such a particle at $\bar{x} = \bar{x}_0$ on $S^2$, becomes, on $S_F^2$:

$$\langle \psi_\alpha(\bar{x}) \| \psi(\bar{x}_0) \rangle \rightarrow k\sigma_i \langle \zeta_\alpha \| \zeta_\alpha \rangle = k\sigma_i \quad . \tag{4.20}$$

The probability $\left| \langle \psi_\alpha(\bar{x}) \| \psi(\bar{x}_0) \rangle \right|^2$ becomes:

$$\left| \langle \psi_\alpha(\bar{x}) \| \psi(\bar{x}_0) \rangle \right|^2 \rightarrow k^2\sigma_i^2 = 1 \tag{4.21}$$

that is: $p(\xi_1)^2 + p(\xi_2)^2 = 1$, as announced in Eq. (4.9).

## 5. Conclusions

Regarding the classical Holographic Principle, it is not necessary to consider quantum field theories, just consider gravity and quantum mechanics, or rather, geometrical quantization as in the case of Loop Quantum Gravity. But the principle was not yet proven.

We think that the Holographic Principle can only be demonstrated in its quantum version (QHP) at least in the LQG framework, and to do this, DQFT (in particular, fermionic DQFT) must be introduced.

Indeed, the QFT is only a starting point, albeit an indispensable one, because in the end QFT is deprived of its infinite degrees of freedom, and only its quantum-computational content remains, that is the quantum information that was hidden in the original theory. The quantum information released by the quantum fields gives rise to maximally entangled qubit states [27].

In Section 3, doubling the fermionic degrees of freedom is done at finite volume $V$ (the volume of the sphere $S^2$). For $V$ which tends to infinity, one will obtain infinite representations of the CCR which are unitarily inequivalent to each other. In this limit the boundary surface becomes a flat screen to infinity, which encodes quantum information entangled with the environment. The relevant consequence is that non-unitarity does not allow disentanglement [39], and this produces a very robust entanglement.

A holographic duality (e.g., the AdS/CFT correspondence) is in general a duality between a theory with gravity and a theory without gravity, which live in the bulk and at the boundary of a space-time manifold respectively.

In this paper, we have proposed a holographic duality between fermionic DQFT in the bulk, the latter having a discrete geometry through spin networks, and quantum gravity (an empty quantum space at the Planck scale) which encodes quantum information at the boundary. Such an empty quantum space endowed with quantum information is a minimal model of quantum gravity, also called "computational LQG" (or CLQG) [13]. It should be noted that on the boundary surface there is no longer any trace of QFT, except in the form of quantum information, the latter being that released by quantum fields.

In the QHP, the entropy bound is different from the usual one because the degrees of freedom encoded by the boundary surface are not Booleans, but they are qubits. It follows that the number of degrees of freedom is no longer directly proportional to the area of the boundary surface, but has a logarithmic growth, obviously much slower. This means that a very small number of qubits is sufficient to encode, on the boundary surface, all the information entropy stored by the bulk. Or, in other words, a quantum computation performed by the black hole (for black holes such as quantum computers, see for example [40] [41] [42]) is much faster. Or again, to simulate a black hole you would have to use a quantum algorithm that is exponentially faster than the classical analogue.

To conclude, for now we see quantum gravity as quantum geometry at the Planck scale, supplemented by quantum information theory. That is, quantum gravity appears as a purely theory of space-time at the Planck scale, where all other interactions have released their quantum information. It must be said that our conclusion does not depend on the commutativity or not of the spatial coordinates: the non-commutative example of the fuzzy sphere is just another way of describing the same result.

It is generally believed that the two completely separate theories of General Relativity and QFT must merge at the Planck scale, but the problem is how. From our results, a direct unification of gravity with other interactions seems impossible. Perhaps gravity can only merge with the quantum information "extracted" from QFT at the Planck scale.

In fact, at the Planck scale, the quantum aspects of gravity become relevant, in the sense that the quantum fluctuations of the metric are comparable with the metric itself [43] [44]. Then, the energy scale of Quantum Gravity will be about $10^{19}$ GeV, while in the Standard Model (SM) the energy scales range from zero to $10^{12}$ eV. We have seen that when one tries to consider QFT at the Planck scale, the original theory ceases to exist and in its place only the quantum information that was contained in it remains. So at the Planck scale we only have quantum gravity (quantum space) encoding quantum information (released by quantum fields).

Conversely, if we look at quantum gravity on the SM scale, the quantum aspects of gravity vanish. We are left with only quantum fields, but at that scale their quantum information is unattainable. Therefore, the two different scales prevent the unification, at least in the way normally conceived, of quantum gravity with the other interactions.

This does not seem like a serious problem to us, because in effect quantum gravity has taken over, revealed and encoded all the hidden quantum information hold by quantum fields.

It is indeed from the "It from qubit" [14] proposal based on Wheeler's "It from bit" [32] that it appears possible to consider a quantum-computational space. In this paper, we have also shown that the quantum information encoded by this space is that released by quantum fields at the Planck scale. At this scale, space is therefore empty; it is a quantum vacuum, or rather a collection of infinite unitarily inequivalent vacua, due to the u.i.r. of the CCR in QFT.

In [42] the hypothesis was made that such a type of space could compute the fundamental laws of physics, in agreement with Lloyd's view [40] [41].

A quantum computation is a unitary operation; therefore the quantum-computational space at the Planck scale can only perform unitary operations. But a certain quantum computation happens in a very precise empty quantum space, one of the infinite unitarily inequivalent ones. A different quantum computation will take place in another vacuum unitarily inequivalent to the one considered above. If we consider the non-unitarity of Hawking's black holes evaporation [45], it can be admitted as tunnelling between unitarily inequivalent vacua.

## Acknowledgments

I am very grateful to Giuseppe Vitiello for interesting discussions and useful comments.